\documentclass[twocolumn,showpacs,preprintnumbers]{revtex4}
\usepackage{amssymb}
\usepackage{amsmath}
\usepackage{dcolumn}
\usepackage{bm}
\usepackage{graphicx}

\begin{document}

\title{High-fidelity quantum memory using nitrogen-vacancy center ensemble
for hybrid quantum computation}
\author{W. L. Yang$^{1}$}
\email{ywl@wipm.ac.cn}
\author{Z. Q. Yin$^{2}$, Y. Hu$^{3}$}
\author{M. Feng$^{1}$}
\email{mangfeng@wipm.ac.cn}
\author{J. F. Du$^{4}$}
\email{djf@ustc.edu.cn}
\affiliation{$^{1}$State Key Laboratory of Magnetic Resonance and Atomic and Molecular
Physics, Wuhan Institute of Physics and Mathematics, Chinese Academy of
Sciences, and Wuhan National Laboratory for Optoelectronics, Wuhan 430071,
China}
\affiliation{$^{2}$Key Laboratory of Quantum Information, University of Science and
Technology of China, Chinese Academy of Sciences, Hefei 230026, China}
\affiliation{$^{3}$Department of physics, Huazhong University of Science and Technology,
Wuhan 430074, China}
\affiliation{$^{4}$Hefei National Laboratory for Physics Sciences at Microscale and
Department of Modern Physics, University of Science and Technology of China,
Hefei, 230026, China}

\begin{abstract}
We study a hybrid quantum computing system using nitrogen-vacancy center
ensemble (NVE) as quantum memory, current-biased Josephson junction (CBJJ)
superconducting qubit fabricated in a transmission line resonator (TLR) as
quantum computing processor and the microwave photons in TLR as quantum data
bus. The storage process is seriously treated by considering all kinds of
decoherence mechanisms. Such a hybrid quantum device can also be used to
create multi-qubit W states of NVEs through a common CBJJ. The experimental
feasibility and challenge are justified using currently available technology.
\end{abstract}

\maketitle

With experimental progress in fabrication of solid-state systems and in
manipulation of qubits in atomic, molecular and optical systems, a lot of
efforts have been given to explore the possibility of building hybrid
quantum devices \cite{HQD}. Combining the advantages of component systems in
a compatible experimental setup, hybrid quantum device may allow us to focus
on the quantum coherent interface between different kinds of qubits.
Examples include storage of quantum information from fragile qubits, e.g.,
superconducting qubit \cite{SC}, to long-lived quantum memories such as
ultracold $^{87}$Rb atomic ensemble \cite{Rb1} and polar molecular ensemble
\cite{Pol1,Pol2}. To achieve appreciable coupling strengths, these systems
often resort to large electric-dipole interactions \cite{Rb1,Pol1}, e.g.,
employing optically excited Rydberg states \cite{Rb1}.

\begin{figure}[tbp]
\includegraphics[width=7 cm]{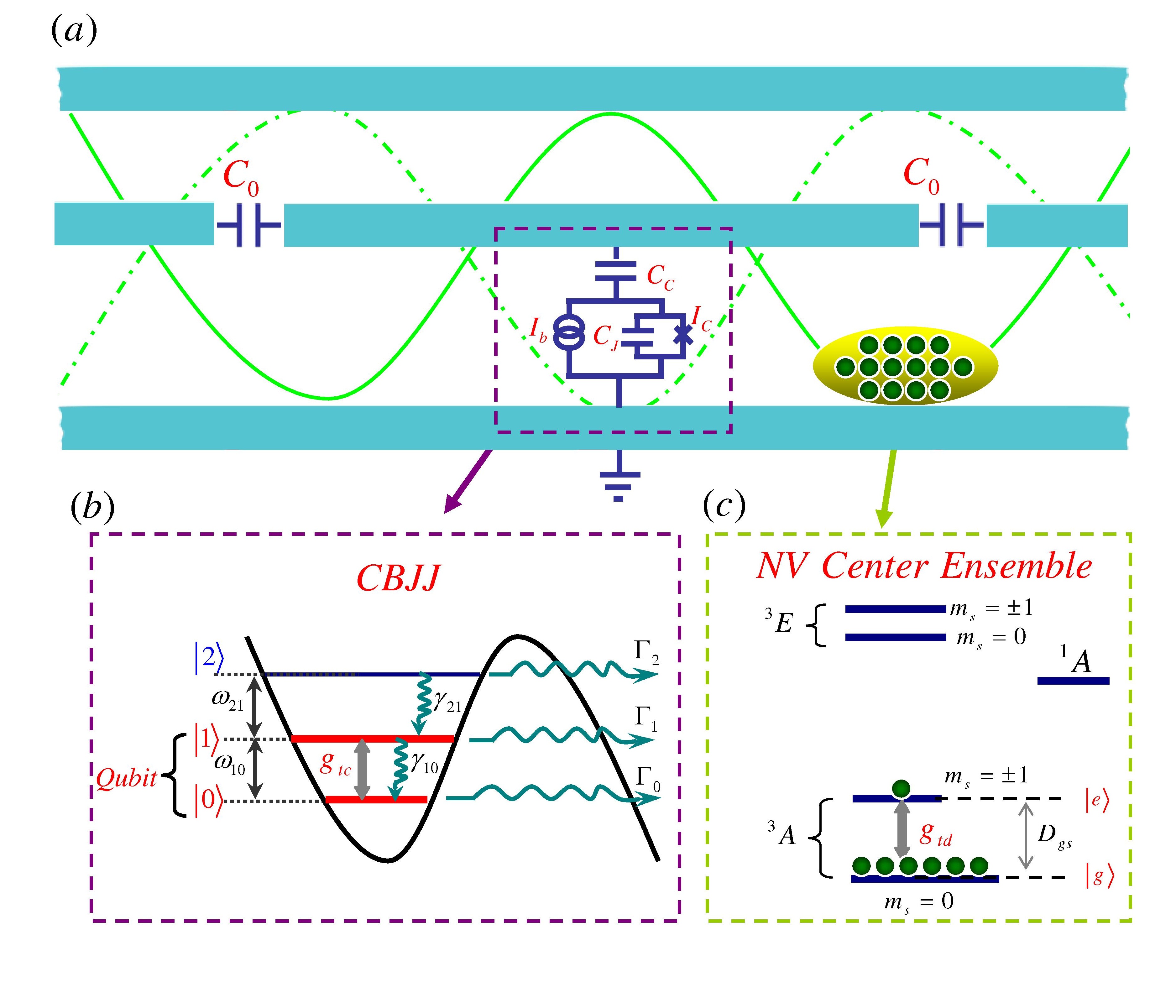}
\caption{(Color online) (a) A NVE and a CBJJ are coupled via a common
quantized field of the TLR acting as a quantum data bus. The CBJJ is
fabricated at the antinodes of the electric field of the full-wave mode of
the TLR. (b) Level scheme of a CBJJ qubit. When the bias current $I_{b}$ is
driven close to the critical bias $I_{c}$, there exist only a few bound
states $\left\vert n\right\rangle _{C}$ with energy $E_{n}$ in each
washboard well \protect\cite{Mar}. (c) Level structure of a NVE where the
electronic ground and first excited states are electron spin triplet states
(S=1), and $D_{gs}/2\protect\pi =2.87$ GHz is the zero-field splitting
between the lowest energy $m_{s}=0$ sublevel and the $m_{s}=\pm 1$
sublevels. }
\end{figure}

However, magnetic-dipole interactions may be more desirable due to
sufficiently long coherence times achieved in systems with spin states
storing quantum information \cite{Marc}. Besides, electronic spin degrees of
freedom may provide excellent quantum memory owing to their weak magnetic
interaction with the environment. For example, the nitrogen-vacancy (NV)
centers in diamond have significant electronic spin lifetime, narrow-band
optical transitions, as well as the possibility of coherent manipulation at
room temperature \cite{Chi1,Chi2}. Several recent highlights include two
parallel experiments demonstrating strong magnetic couplings ($\sim $ 10
MHz) of a transmission line resonator (TLR) to an ensemble of $10^{11}\sim
10^{13}$ NV centers \cite{Kubo}, and to $N$ substitution (P1) centers \cite%
{Schu}, where the strong spin-field coupling benefits from the enhancement
of the coupling strength by a factor $\sqrt{N}$ for an ensemble consisting
of $N$ qubits. It implies the possibility of using electronic spin ensemble
to construct a good quantum memory for superconducting qubits.

In this work, we study a hybrid quantum device which uses nitrogen-vacancy
center ensemble (NVE) as a spin-based quantum memory and uses current-biased
Josephson junction (CBJJ) phase qubit \cite{Yu,Mar,C1,C2,Yang} fabricated in
a TLR \cite{Bl} as a quantum computing processor, which implements rapid
quantum logic gates. The spin qubits are coupled to quantized modes of TLR
through an effectively enhanced magnetic-dipole interaction, which enables
high-quality storage of arbitrary quantum states of a CBJJ. Considering
decoherence in experimentally available systems, we show the feasibility to
achieve a high-fidelity quantum memory with either resonant interaction (RI)
or dispersive interaction (DI) by modulating the external parameters of the
CBJJ to independently tune both the CBJJ's transition frequency and CBJJ-TLR
coupling strength. We also discuss a potential application of such a hybrid
quantum device for preparing multi-qubit W states of NVEs.

As illustrated in Fig. 1(a), the hybrid quantum device we study is the
CBJJ-TLR-NVE system governed by $H_{tot}=H_{C}+H_{T}+H_{D}+H_{TC}+H_{TD}$.
The employed CBJJ is designed using Josephson junctions to make low loss
nonlinear oscillator so that one of the key elements of the CBJJ, i.e.,
sufficiently large anharmonicity, could be obtained for preventing qubit
operations from exciting other transitions in these energy levels. So the
Hamiltonian of CBJJ can be simply modeled as $H_{C}=\frac{\hbar }{2}\omega
_{10}\sigma ^{z}$, where $\sigma ^{z}=\left\vert 1\right\rangle
_{C}\left\langle 1\right\vert -\left\vert 0\right\rangle _{C}\left\langle
0\right\vert $ is the Pauli spin operator, and $\hbar \omega _{10}$ is the
lowest energy-level spacing of the CBJJ. With the energy of $\left\vert
0\right\rangle _{C}$ to be energy zero point, the energy-level spacings can
be expressed as $\omega _{10}\simeq 0.9\omega _{p}$ and $\omega _{21}\simeq
0.81\omega _{p}$ with $\omega _{p}=\sqrt[4]{(2-2I_{b}/I_{c})(2\pi I_{c}/\Phi
_{0}C_{J})^{2}}$ the plasma oscillation frequency at the bottom of the well
\cite{Mar}, the bias current $I_{b}$, the critical current $I_{c}$, the
junction capacitance $C_{J}$, and the flux quantum $\Phi _{0}=h/2e$.

The Hamiltonian of the microwave-driven TLR (with length $L$, inductance $%
F_{t}$, capacitance $C_{t}$, wiring capacitors $C_{0}$, and coupling
capacitors $C_{c}$) is $H_{T}=\hbar \omega _{c}a^{\dagger }a$ where $a$ $%
(a^{\dagger })$ is the annihilation (creation) operator of the full-wave
mode, and the frequency of this mode is slightly renormalized by the wiring
and coupling capacitor as $\omega _{c}\approx 2\pi (1-\varepsilon
_{1}-\varepsilon _{2})/(\sqrt{F_{t}C_{t}})$ with $\varepsilon _{1}=$ $%
2C_{0}/C_{t}$ and $\varepsilon _{2}=$ $C_{c}/C_{t}$. The voltage and current
distribute inside the TLR as
\begin{eqnarray}
V_{tlr}(x) &=&\sqrt{\hbar \omega _{c}/C_{t}}(a^{\dagger }+a)\cos (kx+\delta
_{0}),  \notag \\
I_{tlr}(x) &=&-i\sqrt{\hbar \omega _{c}/F_{t}}(a-a^{\dagger })\sin
(kx+\delta _{0}),
\end{eqnarray}%
where $k=2\pi /L$, and the small phase $\delta _{0}$ meets the condition $%
\tan $ $\delta _{0}=2\pi \varepsilon _{2}$. The TLR can be capacitively
coupled to the CBJJ by the coupling capacitors $C_{c}$ and junction
capacitance $C_{J}$. This resonator acts as the channel to control, couple,
and read out the states of the qubit \cite{Gam}. In this circuit quantum
electrodynamic architecture, the CBJJ-TLR interaction Hamiltonian can be
written as $H_{TC}=\hbar g_{tc}(\sigma ^{+}a+\sigma ^{-}a^{\dagger })$ with $%
\sigma ^{+}=\left\vert 1\right\rangle _{C}\left\langle 0\right\vert $ ($%
\sigma ^{-}=\left\vert 0\right\rangle _{C}\left\langle 1\right\vert$) the
CBJJ's rising (lowering) operator. $g_{tc}=[2C_{t}(C_{J}+C_{c})]^{-1/2}%
\omega _{c}C_{c}\cos \delta _{0}$ is a tunable qubit-resonator coupling
strength \cite{C2} due to the flexibility of CBJJ-qubits, allowing access to
a wide range of tunable experimental parameters for $I_{b}$ and $C_{J}$. A
similar system was presented in Ref. \cite{Yang} for the realization of
entanglement between two distant NVEs.

The Hamiltonian of a NVE reads $H_{D}=$ $\frac{\hbar }{2}\omega _{eg}S^{z}$,
where $S^{l}$ $(l=z,\pm )$ $=(1/\sqrt{N})\sum_{i}^{N}\tau _{i}^{l}$ is the
collective spin operator for the NVE with $\tau ^{z}=\left\vert
e\right\rangle \left\langle e\right\vert -\left\vert g\right\rangle
\left\langle g\right\vert $, $\tau ^{+}=\left\vert e\right\rangle
\left\langle g\right\vert $ and $\tau ^{-}=\left\vert g\right\rangle
\left\langle e\right\vert $. The operator $S^{+}$ can create symmetric Dicke
excitation states $\left\vert n\right\rangle _{D}$, among which the NVE
could be defined in the lowest two states $\left\vert 0\right\rangle
_{D}=\left\vert g_{1}g_{2\cdot \cdot \cdot }g_{N}\right\rangle $ and $%
\left\vert 1\right\rangle _{D}=S^{+}\left\vert 0\right\rangle _{D}=(1/\sqrt{N%
})\sum\nolimits_{k=1}^{N}\left\vert g_{1\cdot \cdot \cdot }e_{k\cdot \cdot
\cdot }g_{N}\right\rangle $. All the spins in NVE interact symmetrically
with a single mode of electromagnetic field because the mode wavelength is
larger than the spatial dimension of the NVE if the spin ensemble is placed
near the TLR's field antinode. So the NVE can be coupled to the TLR by
magnetic-dipole coupling with the corresponding Hamiltonian $H_{TD}=\hbar
g_{td}(S^{+}a+S^{-}a^{\dagger })$, which is a Jaynes-Cummings-type
interaction with $g_{_{td}}=\sqrt{N}g_{s}$, and $g_{s}$ being the single NV
vacuum Rabi frequency.

\begin{figure}[tbph]
\centering\includegraphics[width=6.9 cm]{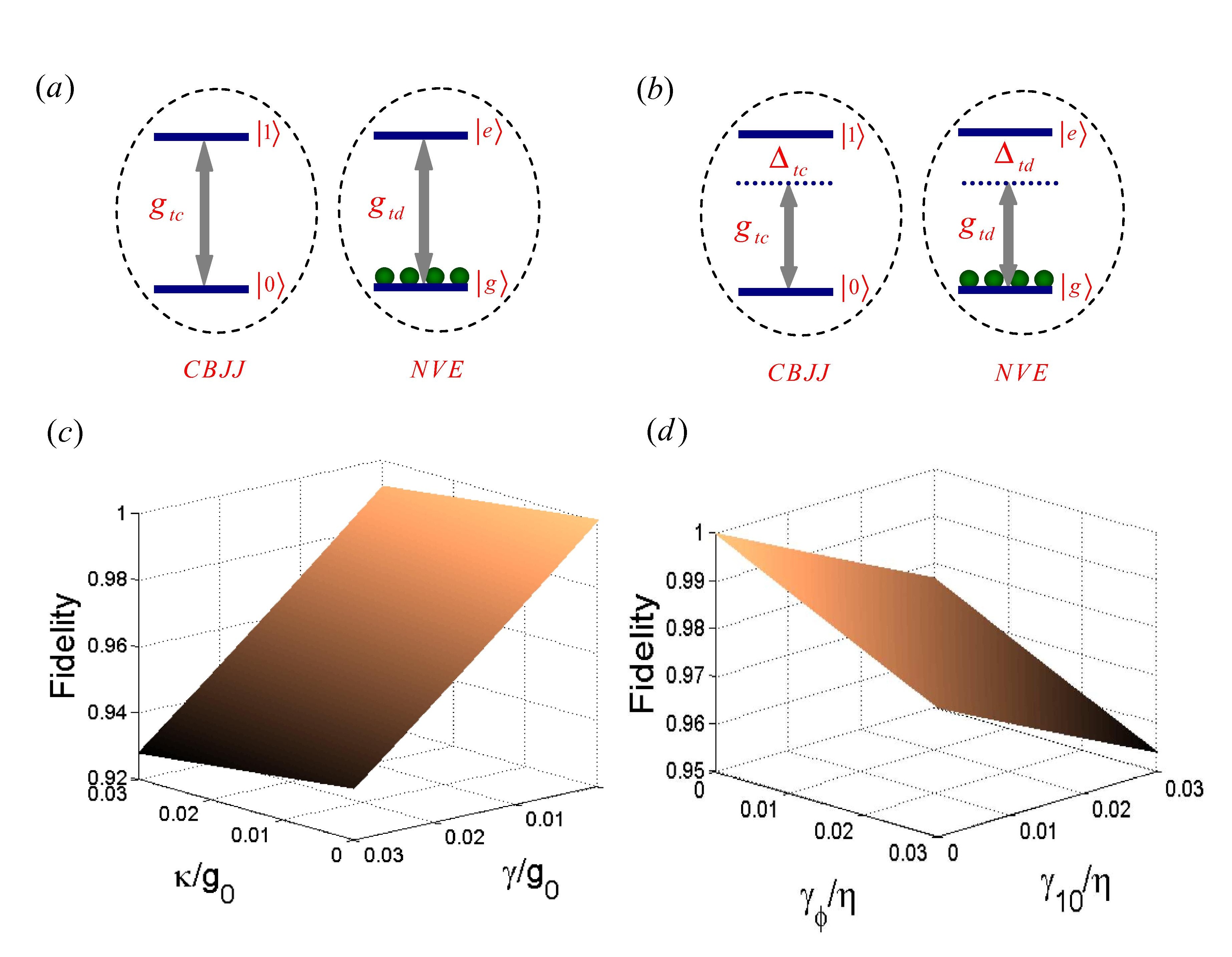} \caption{(Color
online) (a) Level Scheme of the CBJJ and NVE coupled to the TLR in
resonant case. (b) Level Scheme of the CBJJ and NVE coupled to the
TLR in dispersive case. (c) The fidelity of the quantum state
transfer versus the dimensionless TLR decay rate $\protect\kappa/g_{0}$ and
CBJJ decay rate $\protect
\gamma/g_{0}$ in the case of $\protect\alpha =\protect\beta =1/\protect\sqrt{2}$,
where we have set $\protect\gamma =\protect\gamma _{\protect\phi }=\protect%
\gamma _{10}=\Gamma _{1}$, and $t_{R}=\protect\pi /\protect\sqrt{2}g_{0}$.
(d) The fidelity of the quantum state transfer versus the dimensionless $\protect\gamma _{10}/\eta$
and $\protect\gamma _{\protect\phi}/\eta$ in the case of $\protect\alpha =%
\protect\beta =1/\protect\sqrt{2}$, where we have set $\Gamma _{1}=\protect%
\gamma _{10}$, and $t_{D}=\protect\pi /2\protect\eta $. }
\end{figure}

In the following, we investigate how to realize the NVE-based quantum memory
using RI or DI. The Hamiltonian for this hybrid system is given in units of $%
\hbar =1$ by,
\begin{eqnarray}
H_{tot} &=&\frac{\omega _{10}}{2}\sigma ^{z}+\omega _{c}a^{\dagger }a+\frac{%
\omega _{eg}}{2}S^{z}  \notag \\
&&+g_{tc}(\sigma ^{+}a+\sigma ^{-}a^{\dagger
})+g_{td}(S^{+}a+S^{-}a^{\dagger }).
\end{eqnarray}%
In the RI case, as shown in Fig. 2(a), we should tune the frequency of the
CBJJ to be resonant with the TLR and NVE, namely, $\omega _{10}=\omega
_{c}=\omega _{eg}$, and for simplicity we assume $g_{tc}=g_{td}=g_{_{0}}$.
In the frame rotating with the TLR frequency $\omega _{c}$, Eq. (2) becomes $%
H_{R}=g_{0}[(\sigma ^{+}a+\sigma ^{-}a^{\dagger })+(S^{+}a+S^{-}a^{\dagger
})]$, by which an arbitrary state of CBJJ $\left\vert \Psi _{C}\right\rangle
=\alpha \left\vert 0\right\rangle _{C}+\beta \left\vert 1\right\rangle _{C}$
is first converted into a superposition state of the microwave photon in the
TLR, and then transferred to the NVE as $\left\vert \Psi _{D}\right\rangle
=\alpha \left\vert 0\right\rangle _{D}+\beta \left\vert 1\right\rangle _{D}$
with $\alpha $ and $\beta $ the normalized complex numbers. Once the single
microwave photon is absorbed by the NVE, our goal is achieved, where the
total operation time is $t_{R}=(2k+1)\pi /\sqrt{2}g_{_{0}}$ with $k$ natural
numbers.

In the DI case, as shown in Fig. 2(b), the frequency of TLR\ should be
changed to be detuned from the zero-field splitting of the NV centers by $%
\Delta _{td}=\omega _{eg}-\omega _{c}\gg g_{td}$. Additionally, using fast
control of bias current $I_{b}$ and junction capacitance $C_{J}$ of the
CBJJ, we may tune the frequency $\omega _{10}$ from being resonant with TLR
to a large-detuning case $\Delta _{tc}=\omega _{10}-\omega _{c}\gg g_{tc}$.
As a result, both CBJJ and NVE can dispersively interact with the resonator.
In this limit, using the Fr\"{o}hlich's transformation \cite{Fro}, $H_{tot}$
(Eq. (2)) is rewritten as
\begin{eqnarray}
H_{D1} &=&\omega _{c}a^{\dagger }a+\frac{\omega _{10}}{2}\sigma ^{z}+\frac{%
\omega _{eg}}{2}S^{z}+\frac{g_{tc}^{2}}{\Delta _{tc}}(\sigma ^{+}\sigma
^{-}+\sigma ^{z}a^{\dagger }a)  \notag \\
&&+\frac{g_{td}^{2}}{\Delta _{td}}(S^{+}S^{-}+S^{z}a^{\dagger }a)+\eta
(S^{+}\sigma ^{-}+S^{-}\sigma ^{+}),
\end{eqnarray}%
where $\eta =g_{tc}g_{td}(1/2\Delta _{tc}+1/2\Delta _{td})$ is the effective
CBJJ-NVE coupling rate. If the TLR is initially prepared in the vacuum state
$\left\vert 0\right\rangle _{T}$ then the Hamiltonian $H_{D1}$ becomes
\begin{eqnarray}
H_{D2} &=&(\frac{\omega _{10}}{2}+\frac{g_{tc}^{2}}{2\Delta _{tc}})\sigma
^{z}+(\frac{\omega _{eg}}{2}+\frac{g_{td}^{2}}{2\Delta _{td}})S^{z}  \notag
\\
&&+\eta (S^{+}\sigma ^{-}+S^{-}\sigma ^{+}).
\end{eqnarray}%
Assuming $\omega _{10}=\omega _{eg}$ and $g_{tc}^{2}/\Delta
_{tc}=g_{td}^{2}/\Delta _{td}$, we have the effective Hamiltonian in the
interaction picture as $H_{D}=\eta (S^{+}\sigma ^{-}+S^{-}\sigma ^{+})$,
which implies that the photon-assisted transitions cannot happen in
practice, but the CBJJ-NVE interaction is effectively induced. Different
from the RI case, the DI case is usually mediated by the exchange of virtual
photons rather than real photons, which could effectively avoid the
TLR-induced loss. Therefore the DI method does not require exact time
control for the coupling of the photon qubits to the CBJJ and to the NVE. In
this way, we can also achieve our goal of NVE-based quantum memory, i.e., $%
(\alpha \left\vert 0\right\rangle _{C}+\beta \left\vert 1\right\rangle
_{C})\left\vert 0\right\rangle _{D}\rightarrow \left\vert 0\right\rangle
_{C}(\alpha \left\vert 0\right\rangle _{D}+\beta \left\vert 1\right\rangle
_{D})$, where the degree of freedom of the TLR mode has been eliminated, and
the total operation time is $t_{D}=(2k+1)\pi /2\eta $.

\begin{figure}[tbph]
\centering\includegraphics[width=8.9 cm]{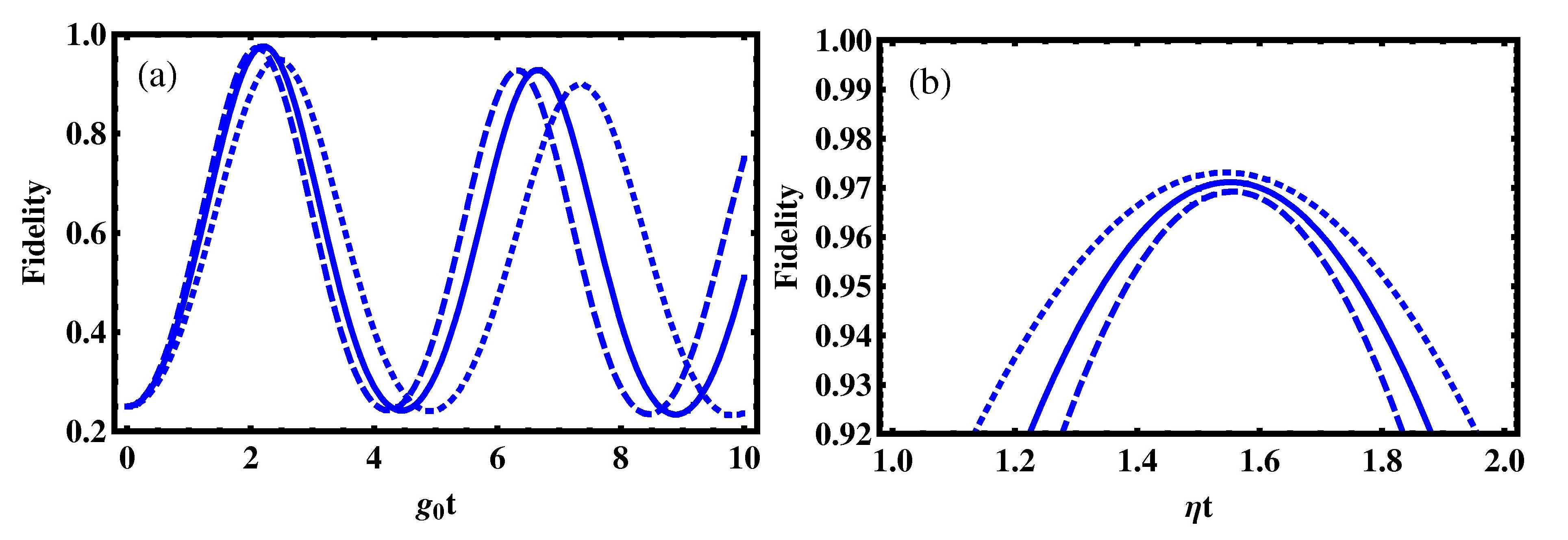}
\caption{(Color online) (a) The fidelity of the quantum state transfer
versus the dimensionless time $g_{0}t$ with $\protect\delta =0$ (solid), $%
-0.1$ (dotted), and $0.1$ (dashed) in the case of $\protect\alpha =\protect%
\beta =1/\protect\sqrt{2}$, $\protect\gamma _{\protect\phi }=\Gamma _{1}=%
\protect\gamma _{10}=\protect\kappa =0.01g_{_{0}}$. (b) The fidelity of the
quantum state transfer versus the dimensionless time $\protect\eta t$ with $%
\protect\alpha =\protect\sqrt{1/2}$ (solid), $\protect\alpha =\protect\sqrt{%
1/3}$ (dashed), and $\protect\alpha =\protect\sqrt{2/3}$ (dotted), where $%
\protect\gamma _{\protect\phi }=\Gamma _{1}=0.015\protect\eta $,\emph{\ }and%
\emph{\ }$\protect\gamma _{10}=0.03\protect\eta $.}
\end{figure}

Taking account of the decoherence effects, we simulate the dynamics of the
transfer process by integrating the full phenomenological quantum master
equation
\begin{equation}
\dot{\rho}=-i[H_{R},\rho ]+\frac{\kappa }{2}D[a]\rho +\frac{\gamma
_{10}+\Gamma _{1}}{2}D[\sigma ^{-}]\rho +\frac{\gamma _{\phi }}{2}D[\sigma
_{z}]\rho ,
\end{equation}%
where $D[A]\rho =2A\rho A^{+}-A^{+}A\rho -\rho A^{+}A$, and $\kappa $ is the
TLR decay rate. $\gamma _{10}$ and $\Gamma _{1}$ are the spontaneous
emission rate and quantum tunneling rate of the level $\left\vert
1\right\rangle _{C}$ of the CBJJ, respectively, and $\gamma _{\phi }$ is the
pure dephasing rate of the CBJJ. We characterize the transfer process for
some given initial states of the CBJJ by the conditional fidelity of the
quantum state according to the expression $F=\left\langle
\Psi_{D}\right\vert \rho (t)\left\vert \Psi _{D}\right\rangle$, where $%
\left\vert \Psi _{D}\right\rangle $ is the target state $\alpha \left\vert
0\right\rangle _{D}+\beta \left\vert 1\right\rangle _{D}$ to be stored in
the NVE. In Fig. 2(c), we plot the fidelity $F$ as a function of the TLR
decay rate $\kappa$ and CBJJ decay rate $\gamma $ in the RI case. One can
find that high fidelity could be achieved in the weak decoherence case, and
the influence from CBJJ seems more serious than that from TLR. In the DI
case, the TLR-induced loss could be effectively suppressed due to the
large-detuning. As shown in Fig. 2(d), we can also obtain high fidelity if
the CBJJ decay rate $\gamma /\eta $ is within the region $[0,0.01]$.

In the RI case, we assume $g_{tc}=g_{0}$ and introduce the parameter
$\delta =(g_{td}-g_{tc})/g_{0}$. In real experiments, $\delta $ is
inevitably non-zero, which leads to errors. But this imperfection is
nearly negligible (See Fig. 3(a)) if $\delta $ is within the region
$[-0.1,0.1]$. Moreover, the state transmission process accelerates
for $\delta >0$, but decelerates in the case of $\delta <0$. In the
DI case, the fidelity $F$ is plotted in Fig. 3(b) against the
operating time for different initial states of the CBJJ, where the
nearly perfect state transfer with the fidelity close to 0.97 is
available if the parameters are chosen as above, and the fidelity
maximum is nearly unchanged with different initial states of the
CBJJ. Noted that the decoherence effects resulted from the
imperfect spin polarization of the NVE could also reduce the fidelity.
Nevertheless, experimental progress has demonstrated the
possibility to suppress the spin depolarization by using suffciently
long optical pulse \cite{JiangL}.

\begin{figure}[tbph]
\centering\includegraphics[width=5.8 cm]{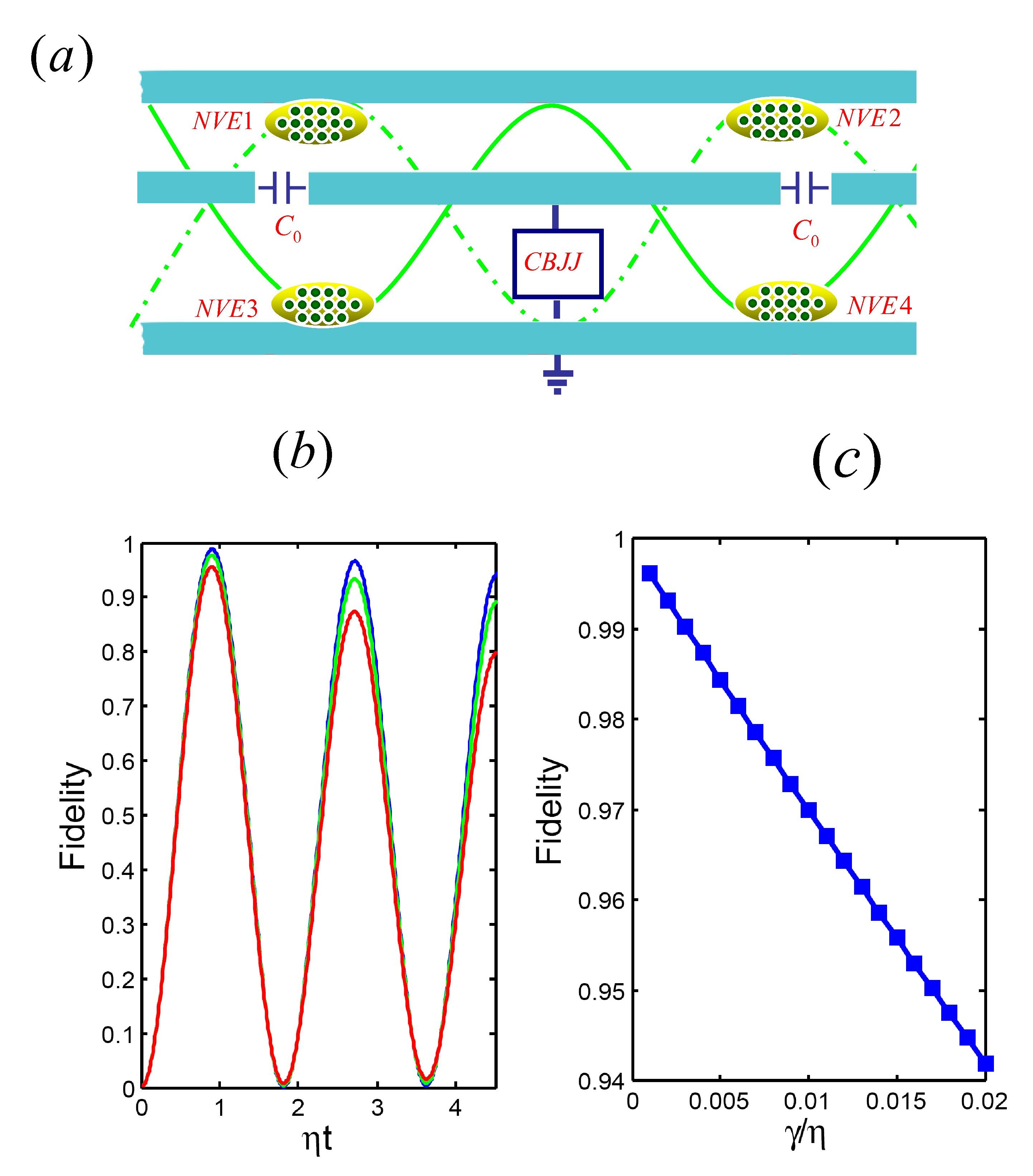} \caption{(Color
online) (a) Schematic of the system consisting of many NVEs coupled
to the TLR, where the CBJJ acts as a tunable coupler. (b) The
fidelity of the state $\left\vert W\right\rangle _{3}$ versus
$\protect\eta t$, where the blue
(upper), green (middle) and red (bottom) curves correspond to $\protect%
\gamma $ = $\protect\eta /50$, $\protect\eta /100$ and $\protect\eta /200$,
respectively, and we have set $\protect\gamma =\protect\gamma _{\protect\phi %
}=\protect\gamma _{10}=\Gamma _{1}$. (c) The fidelity of the state $%
\left\vert W\right\rangle _{3}$ versus the dimensionless parameter $\protect%
\gamma /\protect\eta $, where the gating time is $\protect\pi /2\protect%
\sqrt{3}\protect\eta $, and $\protect\gamma =\protect\gamma _{\protect\phi }=%
\protect\gamma _{10}=\Gamma _{1}$.}
\end{figure}

We survey the relevant experimental parameters. The two methods above
require different conditions in implementation. For example, in the RI case,
the TLR with the inductance $F_{t}=60.7$ nH and the capacitance $C_{t}=2$ pF
leads to a full wave frequency $\omega _{c}/2\pi =2.87$ GHz. The CBJJ's
parameters should be tuned to $C_{J}=71.5$ pF, $C_{c}=60$ fF, $I_{c}=67$ $%
\mu $A, and $I_{b}/I_{c}\approx 0.99$, which make $\omega _{10}=\omega _{c}$
and yield the CBJJ-TLR coupling rate $g_{tc}/2\pi =10$ MHz. In the DI case,
we may choose $\omega _{c}/2\pi =2.62$ GHz and the detuning from the CBJJ
and NVE by $\Delta _{tc}/2\pi =\Delta _{td}/2\pi =250\quad $ MHz. As a
result, in order to obtain a considerable CBJJ-NVE coupling rate $\eta /2\pi
=g_{tc}g_{td}(1/2\Delta _{tc}+1/2\Delta _{td})=10$ MHz in the case of $%
g_{tc}/2\pi =g_{td}/2\pi =50$ MHz, we have to tune the CBJJ's parameters to $%
C_{J}=2.3$ pF, $I_{c}=2.177\mu $A, and $I_{b}/I_{c}\approx 0.99$. The
dissipation parameters $\gamma_{10},\gamma_{\varphi},\Gamma$, and $\kappa$
are on the order of hundreds of kHz \cite{Yu,Mar,Ne}. Therefore, both $g$
and $\eta $ are two orders higher than the dissipation rates, which makes
reliable quantum memory feasible. Finally, let us go back to the two-level
approximation we have made for the CBJJ-qubit. The leakage probability of
quantum state population from the subspace $\{\left\vert 0\right\rangle
_{C}, \left\vert 1\right\rangle _{C}\}$ to $\left\vert 2\right\rangle _{C}$
can be estimated as $P\sim O[g_{tc}^{2}/(g_{tc}^{2}+\Xi ^{2})]\sim 10^{-3}$
with the level separation $\Xi =\left\vert \omega _{21}-\omega
_{10}\right\vert =\omega_{10}/10$. The small value of $P$ ensures that the
two-level approximation in our scheme is reasonable.

One of the favorable applications of our scheme is to entangle many NVEs by
coupling to a common CBJJ, which actually acts as a tunable coupler to
mediate the virtual excitation of photons. As shown in Fig. 4(a), the
effective Hamiltonian of this system in the DI regime reads $%
H_{M}=\sum\nolimits_{i=1}^{N}\eta _{i}(S_{i}^{+}\sigma ^{-}+S_{i}^{-}\sigma
^{+})$, where $\eta _{i}=$ $g_{tc}^{i}g_{td}^{i}(1/2\Delta
_{tc}^{i}+1/2\Delta _{td}^{i})$, and $N$ is the number of NVEs. For
simplicity, we assume that each NVE is equally coupled to the CBJJ, i.e., $%
\eta _{i}=$ $\eta $. So the state evolution under the Hamiltonian $H_{M}$ is
given by $\left\vert \Psi (t)\right\rangle =\cos (\sqrt{N}\eta t)\left\vert
1\right\rangle _{1}\left\vert 1\right\rangle _{2}\cdot \cdot \left\vert
1\right\rangle _{N}\left\vert 0\right\rangle _{c}+$ $(1/\sqrt{N})\sin (\sqrt{%
N}\eta t)\sum\nolimits_{j=1}^{N}\left\vert 1\right\rangle _{1}\cdot \cdot
\left\vert 0\right\rangle _{j}\cdot \cdot \left\vert 1\right\rangle
_{N}\left\vert 1\right\rangle _{c}$. By choosing $\sqrt{N}\eta t$ = $\pi /2$%
, with the readout result of the CBJJ state to be $\left\vert 1\right\rangle
_{c}$, we obtain the $\left\vert W\right\rangle _{N}$ state of NVEs \cite%
{Deng} if the system is initially prepared in the state $\left\vert \Psi
(0)\right\rangle =\left\vert 1\right\rangle _{1}\left\vert 1\right\rangle
_{2}\cdot \cdot \left\vert 1\right\rangle _{N}\left\vert 0\right\rangle _{c}$
by optically pumping the electron spins in NVEs. The readout of the phase
qubit can be accomplished using single-shot measurements \cite{Ste}. Figs.
4(b) and 4(c) show the fidelity of the state $\left\vert W\right\rangle _{3}$
under the consideration of decoherence by the quantum master equation $\dot{%
\rho} =-i[H_{M}, \rho ]+\frac{\gamma _{10}+\Gamma _{1}}{2}D[\sigma^{-}]\rho +%
\frac{ \gamma _{\phi }}{2}D[\sigma _{z}]\rho$.

In summary, we have put forward a realization of a spin-based quantum memory
for CBJJ with currently available techniques. Modulating the external
parameters of the CBJJ, we have achieved the quantum information transfer
between CBJJ and NVE using RI method or DI method under realistic situation.
We argue that our proposal has immediate practical applications in quantum
memory and hybrid quantum device with currently available techniques.

This work is supported by NNSF of China under Grants No. 10974225, No.
11004226 and No. 11074070.


\begin{thebibliography}{99}
\bibitem{HQD} M. Wallquist \textit{et al.}, Phys. Scr. \textbf{T137}, 014001
(2009); A. Imamo\u{g}lu, Phys. Rev. Lett. \textbf{102}, 083602 (2009).

\bibitem{SC} Y. Makhlin, G. Schon, and A. Shnirman, Rev. Mod. Phys. \textbf{%
73}, 357 (2001); A. Wallraff \textit{et al.}, Nature (London) \textbf{431},
162 (2004).

\bibitem{Rb1} D. Petrosyan \textit{et al.}, Phys. Rev. A \textbf{79},
040304(R) (2009).

\bibitem{Pol1} K. Tordrup and K. M\o lmer, Phys. Rev. A \textbf{77}, 020301
(2008).

\bibitem{Pol2} P. Rabl \textit{et al.}, Phys. Rev. Lett. \textbf{97}, 033003
(2006); P. Rabl and P. Zoller, Phys. Rev. A \textbf{76}, 042308 (2007).

\bibitem{Marc} D. Marcos \textit{et al.}, Phys. Rev. Lett. \textbf{105},
210501 (2010).

\bibitem{Chi1} F. Jelezko \textit{et al.}, Phys. Rev. Lett. \textbf{92},
076401 (2004); L. Childress \textit{et al.}, Science \textbf{314}, 281
(2006); T. Togan \textit{et al.}, Nature (London) \textbf{466}, 730 (2010);
P. Neumman \textit{et al.}, Nat. Phys. \textbf{6}, 249 (2010).

\bibitem{Chi2} F. Shi \textit{et al.}, Phys. Rev. Lett. \textbf{105}, 040504
(2010); W. L. Yang \textit{et al.}, Appl. Phys. Lett. \textbf{96}, 241113
(2010); W. L. Yang \textit{et al.}, New. J. Phys. \textbf{12}, 113039 (2010).

\bibitem{Kubo} Y. Kubo \textit{et al.}, Phys. Rev. Lett. \textbf{105},
140502 (2010).

\bibitem{Schu} D. I. Schuster \textit{et al.}, Phys. Rev. Lett. \textbf{105}%
, 140501 (2010).

\bibitem{Yu} Y. Yu \textit{et al.}, Science \textbf{296}, 889 (2002).

\bibitem{Mar} J. M. Martinis \textit{et al.}, Phys. Rev. Lett. \textbf{89},
117901 (2002).

\bibitem{C1} A. Blais \textit{et al.}, Phys. Rev. Lett. \textbf{90}, 127901
(2003); A. M. Zagoskin \textit{et al.}, Phys. Rev. Lett. \textbf{97}, 077001
(2006).

\bibitem{C2} Y. Hu \textit{et al.}, Phys. Rev. A \textbf{75}, 012314 (2007).

\bibitem{Yang} W. L. Yang \textit{et al.}, Phys. Rev. A \textbf{83}, 022302
(2011).

\bibitem{Bl} A. Blais \textit{et al.}, Phys. Rev. A \textbf{69}, 062320
(2004).

\bibitem{Gam} J. M. Gambetta, A. A. Houck, and A. Blais, Phys. Rev. Lett.
\textbf{106}, 030502 (2011).

\bibitem{Fro} H. Fr\"{o}hlich, Phys. Rev. \textbf{79}, 845 (1950); H. R.
Zhang \textit{et al.}, Phys. Rev. A \textbf{80}, 062308 (2009); Q. Ai
\textit{et al.}, Phys. Rev. A \textbf{78}, 022327 (2008).

\bibitem{JiangL} L. Jiang, \textit{et al.}, Science \textbf{326}, 267
(2009); P. Neumann, \textit{et al.}, Science \textbf{329}, 542 (2010).

\bibitem{Ne} L. Du \textit{et al.}, New. J. Phys. \textbf{12}, 063015
(2010); L. Frunzio \textit{et al.}, IEEE Trans. Appl. Supercond. \textbf{15}%
, 860 (2005).

\bibitem{Deng} Z. J. Deng, M. Feng, and K. L. Gao, Phys. Rev. A \textbf{73},
014302 (2006).

\bibitem{Ste} M. Steffen \textit{et al.}, Phys. Rev. Lett. \textbf{97},
050502 (2006).
\end{thebibliography}
\end{document}